\documentclass[%
reprint,
superscriptaddress,
 amsmath,amssymb,
 aps,
pra,
]{revtex4-2}

\usepackage{graphicx}
\usepackage{dcolumn}
\usepackage{bm}
\usepackage{amssymb}
\usepackage{breqn}
\usepackage{xcolor}
\usepackage{amsmath}
\usepackage{float}
\usepackage{ulem} 

\begin{document}

\title{
Quantum Advantage with Seeded Squeezed Light for Absorption Measurement
}

\author{Fu Li}
\email{fuli@physics.tamu.edu}
\affiliation{
Institute for Quantum Science and Engineering$,$ Texas A\&M  University${,}$  College Station${,}$   TX  77843${,}$  USA
}
\affiliation{
Department of Physics and Astronomy${,}$  Texas  A\&M  University${,}$ College Station${,}$  TX  77843${,}$  USA
}

\author{Tian Li}
\email{tian.li@tamu.edu}
\homepage{T.L. and F.L. contributed equally to this work.}
\affiliation{
Institute for Quantum Science and Engineering$,$ Texas A\&M  University${,}$  College Station${,}$   TX  77843${,}$  USA
}
\affiliation{
Department of Biological and Agricultural Engineering${,}$  Texas A\&M  University${,}$  College Station${,}$ TX 77843${,}$  USA
}

\author{Marlan O. Scully}
\affiliation{
Institute for Quantum Science and Engineering$,$ Texas A\&M  University${,}$  College Station${,}$   TX  77843${,}$  USA
}
\affiliation{
Department of Mechanical and Aerospace Engineering${,}$  Princeton  University${,}$  Princeton${,}$  NJ  08544${,}$ USA
}
\affiliation{
Quantum Optics Laboratory${,}$  Baylor  Research and  Innovation Collaborative${,}$  Waco${,}$ TX  76704${,}$  USA
}

\author{Girish S. Agarwal}
\affiliation{
Institute for Quantum Science and Engineering$,$ Texas A\&M  University${,}$  College Station${,}$   TX  77843${,}$  USA
}
\affiliation{
Department of Physics and Astronomy${,}$  Texas  A\&M  University${,}$ College Station${,}$  TX  77843${,}$  USA
}
\affiliation{
Department of Biological and Agricultural Engineering${,}$  Texas A\&M  University${,}$  College Station${,}$ TX 77843${,}$  USA
}


\begin{abstract}

Absorption measurement is an exceptionally versatile tool for many applications in science and engineering. For absorption measurements using laser beams of light, the sensitivity is theoretically limited by the shot noise due to the fundamental Poisson distribution of photon number in laser radiation. In practice, the shot-noise limit can only be achieved when all other sources of noise are eliminated. \textcolor{black}{Here, we use seeded squeezed light to demonstrate that direct absorption measurement can be performed with sensitivity beyond the shot-noise limit. We present a practically realizable scheme, where intensity squeezed beams are generated by a seeded four-wave mixing process in an atomic rubidium vapor cell. More than 1.2~dB quantum advantage for the measurement sensitivity is obtained at faint absorption levels ($\leq 10\%$). We also present a detailed theoretical analysis to show that the observed quantum advantage when corrected for optical loss would be equivalent to 3~dB. Our experiment demonstrates a direct sub-shot-noise measurement of absorption that requires neither homodyne/lock-in nor logic coincidence detection schemes. It is therefore very applicable in many circumstances where sub-shot-noise-level absorption measurements are highly desirable.}

\end{abstract}




\maketitle



\section{Introduction}

It has been demonstrated that one can improve the sensitivity and precision of many classical measurement techniques using various quantum states of light~\cite{
PhysRevA.44.3266,Ribeiro:97,Hayat:99,Treps940,Giovannetti1330,PhysRevA.77.053807,Plick_2010,Brida:2010lh,nphoton.2012.300,nphoton.2012.346,TAYLOR20161,Whittaker_2017,s41598-017-06545-w,doi:10.1063/5.0009681} \textcolor{black}{(For instance, the experimental work reported in Ref.~\cite{PhysRevA.44.3266} is a sub-shot-noise measurement of an intensity modulation on one of the quantum-correlated twin beams, and the intensity is modulated by adjusting the transmission of the beam from a liquid-crystal cell)}. Most prominently, sub-shot-noise detection of changes in optical phase have been demonstrated in interferometers using quantum light~\cite{Anderson:17,Plick_2010_1,Kolkiran:08,PhysRevLett.107.113603} and have been implemented for gravitational wave detection~\cite{nphys2083}.
Although a straightforward readily attainable approach to achieve desired performances of a classical measurement is to simply increase the photon flux of the probe light to yield a greater signal-to-noise ratio, it has been proven unfeasible whenever one faces limits on the brightness of the optical probes, for instance, in the case where samples can be altered or damaged by the probe light~\cite{TAYLOR20161,doi:10.1111/php.12572}. It is therefore highly desirable to optimize measurement sensitivity with a fixed amount of input photon flux~\cite{TAYLOR20161}. It is also important to note that for measurement schemes where the sensitivity itself varies with parameters of the measured sample it is possible for the sensitivity to be degraded, potentially requiring either prior knowledge about the optical sample or the addition of a feedback servo loop to ensure a sub-shot-noise performance~\cite{nphoton.2010.268,Yonezawa1514,nphoton.2015.139}. 

Since the intensity measurement of an idealized laser fluctuates with a Poisson distribution
, it is therefore used to define the shot-noise limit (SNL) in optical measurements, and it can only be reached in classical experiments once all other sources of noise are removed. For a direct measurement of optical transmission, the number of photons that pass through a sample is used to estimate the sample's absorption $\alpha$, and thus the estimation sensitivity $\Delta \alpha$ is determined by the SNL. \textcolor{black}{One of the most popular approaches that allow for a sub-SNL measurement of an unknown sample's absorption is to use quantum-correlated beams of photons~\cite{Whittaker_2017,s41598-017-06545-w}. 
For practical applications, the reduction of noise between quantum-correlated beams of photons generated with spontaneous parametric down-conversion (SPDC) 
is widely adopted because of the implementation simplicity and the robustness of quantum nature against the introduction of an absorbing sample~\cite{Iskhakov:16}. In particular, such technique has been implemented in the context of imaging, where a charge-coupled-device (CCD) camera is usually employed to acquire sub-SNL measurement in the spatial domain by detecting correlated photons altogether in the same image captured by the camera~\cite{PhysRevA.77.053807,Brida:2010lh,ncomms3426,Samantaray:2017ff,PhysRevA.95.053849,PhysRevA.100.063828}.
With the inclusion of a spatially absorbing sample, it has been shown that correlated photons can be used to suppress noise in imaging objects to a degree that out-performs classical measurement using an equally efficient detection~\cite{Brida:2010lh,Knyazev:19}. Since absorption measurement is the most versatile tool for many applications in spectroscopy, metrology, chemistry and biology, improving the measurement sensitivity is thus indisputably beneficial to both science and engineering communities. 
It is therefore absolutely valuable for experiments to be performed to observe clear quantum advantages that gained by using quantum states of light in absorption measurements.}

\textcolor{black}{Indeed, quantum advantages in absorption measurements have been demonstrated in a series of experiments carried out with photon pairs generated with SPDC in nonlinear crystals, Refs.~\cite{Ribeiro:97,nphoton.2012.300,Whittaker_2017,s41598-017-06545-w} are some prominent examples. It should be noted that SPDC is not the only source of quantum light, another important quantum source is the squeezed light produced with four-wave mixing (FWM) in atomic vapors or optical fibers~\cite{doi:10.1021/acsphotonics.9b00250}. In fact, squeezed light has been extensively studied for its advantage in phase measurement since the early prediction of Caves~\cite{PhysRevD.23.1693}, some prominent experiments are reported in Refs.~\cite{nphys2083,Gupta:18,Anderson:17,Yonezawa1514}. However, there are hardly any experimental demonstrations using squeezed light to achieve sub-SNL absorption measurement \textcolor{black}{since the seminal work done by Polzik~\textit{et al.}~\cite{PhysRevLett.68.3020}}. In this article, we report a practically realizable experimental scheme using squeezed light for direct absorption measurement. We use intensity squeezed beams generated with a seeded FWM process as the source to demonstrate clear quantum advantages over the SNL. Note that Moreau \textit{et al.}~\cite{s41598-017-06545-w} report a quantum advantage of 0.9~dB using SPDC in direct absorption measurement, while we report a higher quantum advantage of more than 1.2~dB for weak absorption levels ($\leq10\%$) as shown below.}

\textcolor{black}{Our experimental scheme is straightforward - a seeded FWM atomic vapor cell together with an electron-multiplying charge-coupled-device (EMCCD) camera comprise the bulk of what is needed to acquire a sub-SNL absorption measurement. Information containing absorption of the sample being measured can be readily obtained by simply integrating the images captured by the EMCCD camera, no homodyne/lock-in or logic coincidence is required. Our scheme therefore is very applicable in many circumstances where sub-SNL absorption measurement is highly desirable. 
We also provide in this article a theoretical model to analyze and gain insights into the experimental observations.}



\section{Results}
\subsection{\textcolor{black}{Theoretical analysis of the quantum advantage for measurement sensitivity}}


\textcolor{black}{Our intensity squeezed light is generated with the FWM process in an atomic $^{85}$Rb vapor cell~\cite{Dowran:18,PhysRevA.95.023803,Anderson:17,Li:17,Pooser:15,Clark:2014vf}. The atomic medium possesses a large third-order electric susceptibility $\chi^{(3)}$, and when appropriately chosen laser light `seeds' the medium, `twin beams', also known as the `probe' and `conjugate' beams, are produced. The theoretical modeling of the twin beams generation in the FWM process is complex, as in the experiment one deals with the probe and conjugate beams of finite bandwidth. In fact, the bandwidth of the twin beams in our scheme is merely $\sim20$~MHz~\cite{Clark:2014vf,Glasser2012a}, which is much narrower compared to what one generates with SPDCs. Therefore, we can recover many of the aspects of our observations in terms of a theoretical model based on an equivalent \textit{single-mode} description of the probe and conjugate beams~\cite{Li:17}. In brief, we use the single-mode approximation and designate $\hat{a}$ and $\hat{b}$ as the mode operators for the probe and conjugate beams respectively, the final operators after detection can therefore be expressed as} 

\begin{equation}
\begin{aligned} 
\begin{split}
\hat{a}_f = \sqrt{\eta_p}\{\sqrt{1-\alpha}[(\text{cosh} r)\hat{a} + e^{i\theta}(\text{sinh} r)\hat{b}^{\dagger}]+i\sqrt{\alpha}\hat{\nu_{\alpha}}\} \\
+ i\sqrt{1-\eta_p}\hat{\nu_p},\\
\end{split}
\\ {\hat{b}^{\dagger}}_f= \sqrt{\eta_c}[(\text{cosh} r)\hat{b}^{\dagger} + e^{-i\theta}(\text{sinh} r)\hat{a}]
-i\sqrt{1-\eta_c}\hat{\nu}^{\dagger}_c,
\end{aligned} 
\end{equation}
\textcolor{black}{where $r$ is the squeezing parameter of the FWM, $\theta$ is the relative phase between the twin beams (approximately, $\theta\cong  2\pi \times 2\nu_{HF} \times L/c$, where 2$\nu_{HF}$ is the frequency difference between the twin beams and $\nu_{HF}$ is the hyperfine splitting in the electronic ground state of $^{85}$Rb shown in Fig.~\ref{Setup}(b), $L$ is the vapor cell length and $c$ is the speed of light), $1-\eta_p$ and $1-\eta_c$ are the optical losses including imperfect detection quantum efficiencies in the probe and conjugate beam paths respectively, $\alpha$ is the absorption we are interested in measuring, and $\hat{\nu_p}$, $\hat{\nu_c}$ and $\hat{\nu_{\alpha}}$ are the vacuum/noise operators. When a coherent state $\vert\beta\rangle$, $\beta=\vert\beta\vert e^{i\phi}$, where $\phi$ is the input phase, seeds mode $a$, and only vacuum fluctuations $|0\rangle$ seed mode $b$, then the input state can be written as $\vert\beta, 0, 0, 0, 0\rangle$, where the third, fourth and fifth zeros are the inputs for the vacuum/noise operators $\hat{\nu_p}$, $\hat{\nu_c}$ and $\hat{\nu_{\alpha}}$ respectively. Although not trivial, it is fairly straightforward to calculate the number operators $\hat {N_a} = \hat{a}^{\dagger}_f\hat{a}_f$ and $\hat{N_b}=\hat{b}^{\dagger}_f\hat{b}_f$ for the probe and conjugate beams after detection. Since the sample is placed in the probe beam, and the conjugate beam is used as a reference, we adopt 
the photon counts difference $\langle \hat{S_{\alpha}} \rangle= \langle \hat {N_a}-\hat {N_b} \rangle$ as our measurement signal. Note that this double-beam approach is commonly implemented in imaging and spectroscopy applications involving weak absorption~\cite{Samantaray:2017ff,Brida:2010lh}, because it enables the cancellation of classical super-Poissonian noise and provides a direct measurement of the absorption by instantaneous comparison with the unperturbed reference beam. The measurement sensitivity,} 

\begin{equation}
\begin{aligned} 
\Delta \alpha= \frac{\sqrt{\langle\Delta^2 \hat{S_{\alpha}}\rangle}}{|\partial_\alpha \langle \hat{S_{\alpha}} \rangle|}, 
\label{DEL}
\end{aligned} 
\end{equation}
\textcolor{black}{can then be readily obtained. In this article we define the quantum advantage as the ratio of the sensitivity enabled by the squeezed light, $\Delta \alpha_{\text{sqz}}$, to the one acquired from the coherent light, $\Delta \alpha_{\text{snl}}$, with the same amount of average photon numbers $\langle N_a \rangle$ and $\langle N_b \rangle$ as the twin beams:} 

\begin{equation}
\begin{aligned} 
\text{Quantum Advantage (dB)} = 10\times\text{log}_{10}\frac{\Delta \alpha_{\text{sqz}}}{\Delta \alpha_{\text{snl}}}.
\label{QD}
\end{aligned} 
\end{equation}

\begin{figure}[t]
    \begin{center}
    \includegraphics[width=\linewidth]{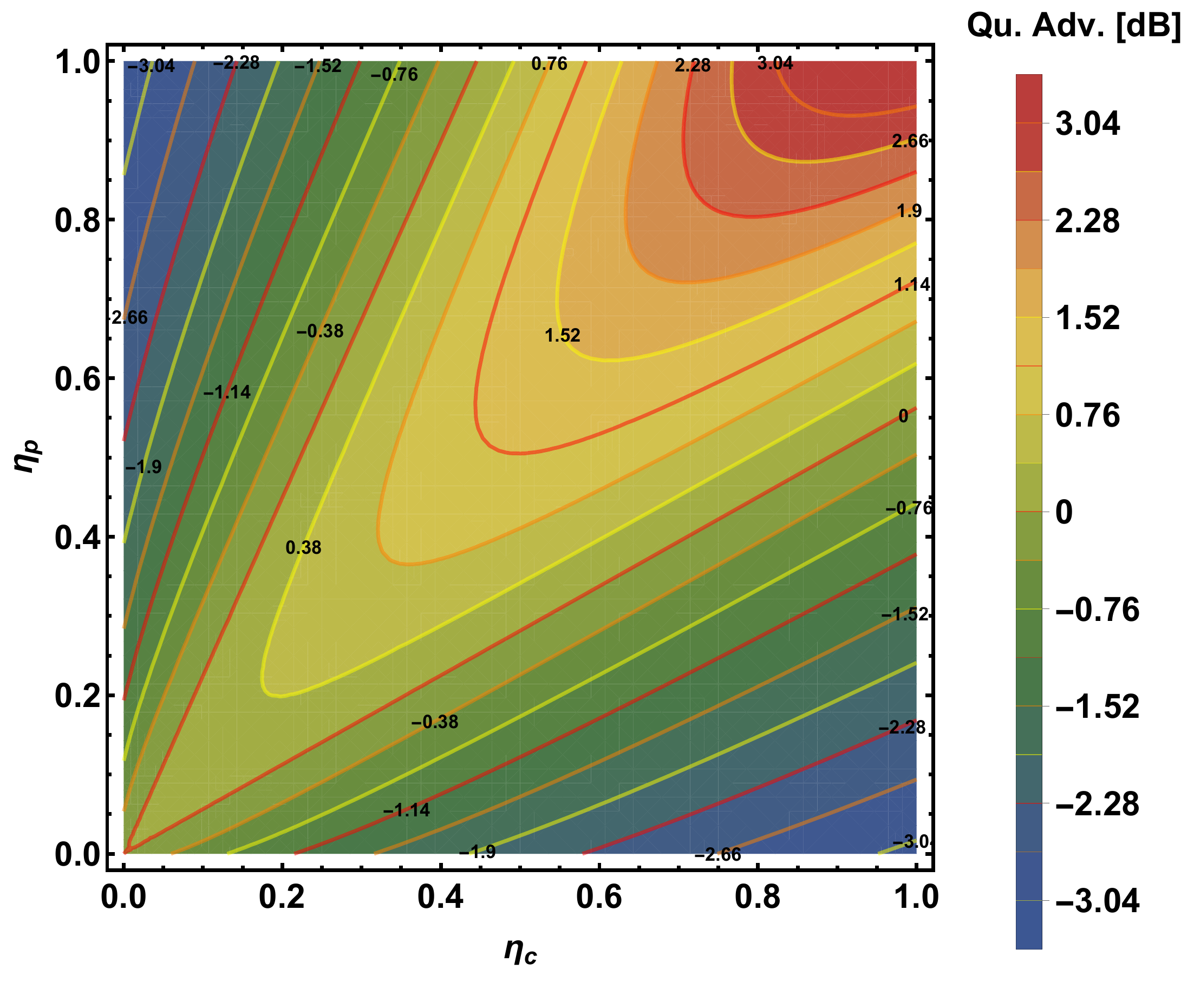}
    \caption{
    \textcolor{black}{Theoretical prediction for the quantum advantage (Qu. Adv.) for absorption $\alpha = 5\%$ as a function of optical transmission in the probe beam path $\eta_p$ and conjugate beam path $\eta_c$. The squeezing parameter $r=1.1$ corresponds to 6.5~dB two-mode squeezing.}
        \label{Theory}}
    \end{center}
\end{figure}

\textcolor{black}{In Fig.~\ref{Theory} we plot the theoretical quantum advantage for absorption $\alpha = 5~\%$ as a function of \textcolor{black}{optical transmission} in the probe beam path $\eta_p$ and conjugate beam path $\eta_c$. The squeezing parameter $r=1.1$, which is calculated from the two-mode squeezing of 6.5~dB~\cite{Li:17} measured by near-perfect photodiodes (see Fig.~\ref{Setup}(c) and Ref.~\cite{Li:20} for further details of the squeezing measurement). It is clear noticeable from the graph that if one could manage to curb the optical loss in both beam paths to be within 10~\%, more than 3~dB quantum advantage for the measurement sensitivity would be readily achievable.}


\subsection{Experimental demonstration of the quantum advantage}

\begin{figure}[htbp]
    \begin{center}
    \includegraphics[width=\linewidth]{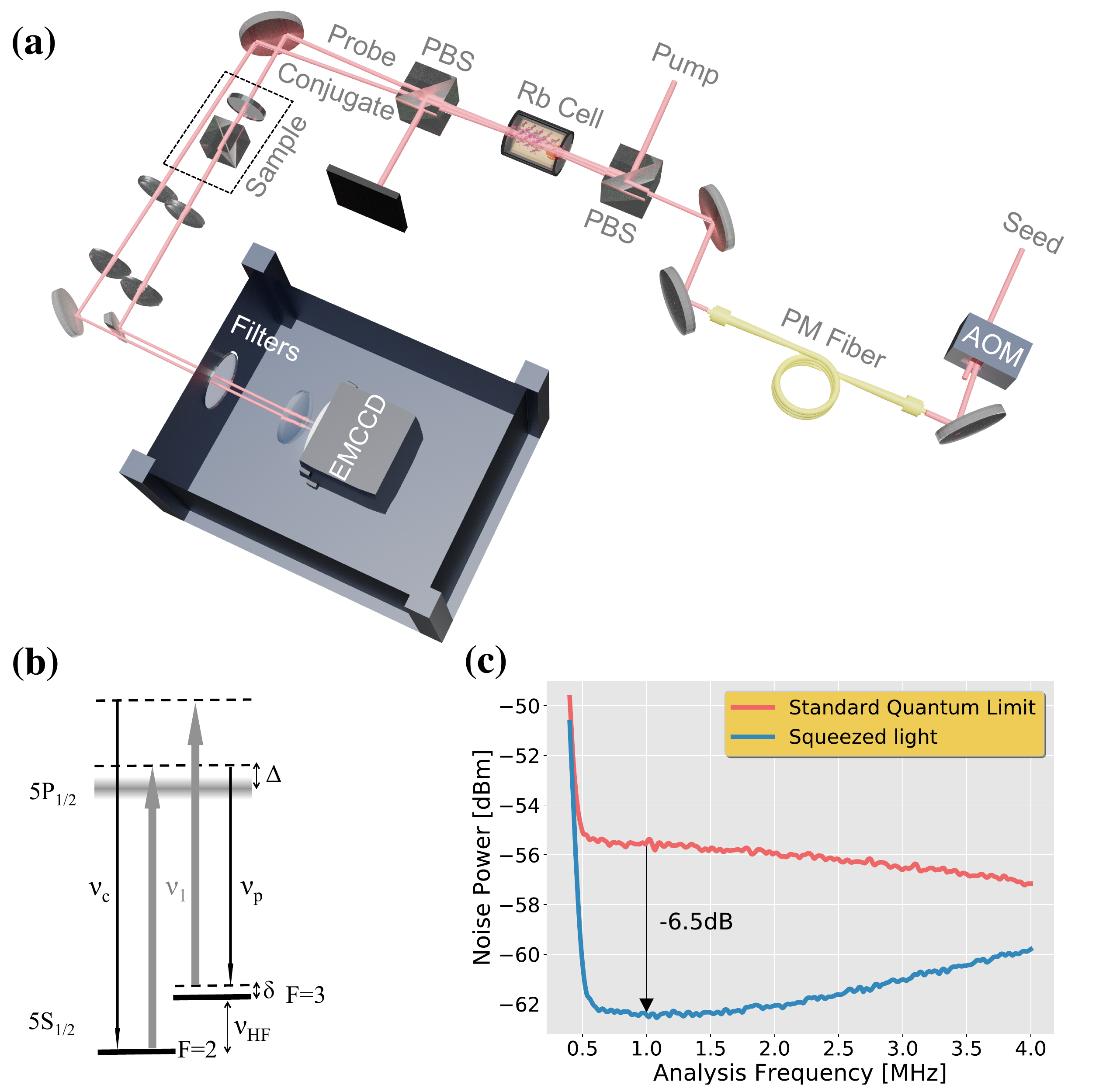}
    \caption{
     (a) Experimental setup in which a seeded $^{85}$Rb vapor cell produces strong quantum-correlated twin beams via FWM. The twin beams are separated from the pump by a $\sim$~$2\times10^5$~$:1$ polarizer after the cell. The probe beam passes through an absorption `sample' (i.e., a combination of a $\lambda/2$ plate and a PBS) while the conjugate beam serves as a reference, before they are focused onto an EMCCD camera. The camera is enclosed in a light-proof box with filters mounted to block ambient light. The AOM in the probe beam path is used to pulse the twin beams with 2~$\mu$s FWHM and duty cycle of $1/12$. PBS: polarizing beam splitter, PM fiber: polarization-maintaining fiber. (b) Level structure of the D1 transition of $^{85}$Rb atom. The optical transitions are arranged in a double-$\Lambda$ configuration, where $\nu_p$, $\nu_c$ and $\nu_1$ stand for probe, conjugate and pump frequencies, respectively, fulfilling $\nu_p$ +  $\nu_c$ =  $2\nu_1$. The width of the excited state in the level diagram represents the Doppler broadened line. $\Delta$ is the one-photon detuning, $\delta$ is the two-photon detuning, and $\nu_{\text{HF}}$ is the hyperfine splitting in the electronic ground state of $^{85}$Rb. \textcolor{black}{(c) Measured intensity-difference noise power spectrum for the squeezed twin beams (blue line) and for the SNL (red line), obtained with a radio frequency spectrum analyzer (with resolution and video bandwidth of 300~kHz and 100~Hz, respectively). A squeezing of 6.5~dB is achieved.} 
        \label{Setup}}
    \end{center}
\end{figure}

\begin{figure*}[htbp]
    \includegraphics[width=0.75\textwidth]{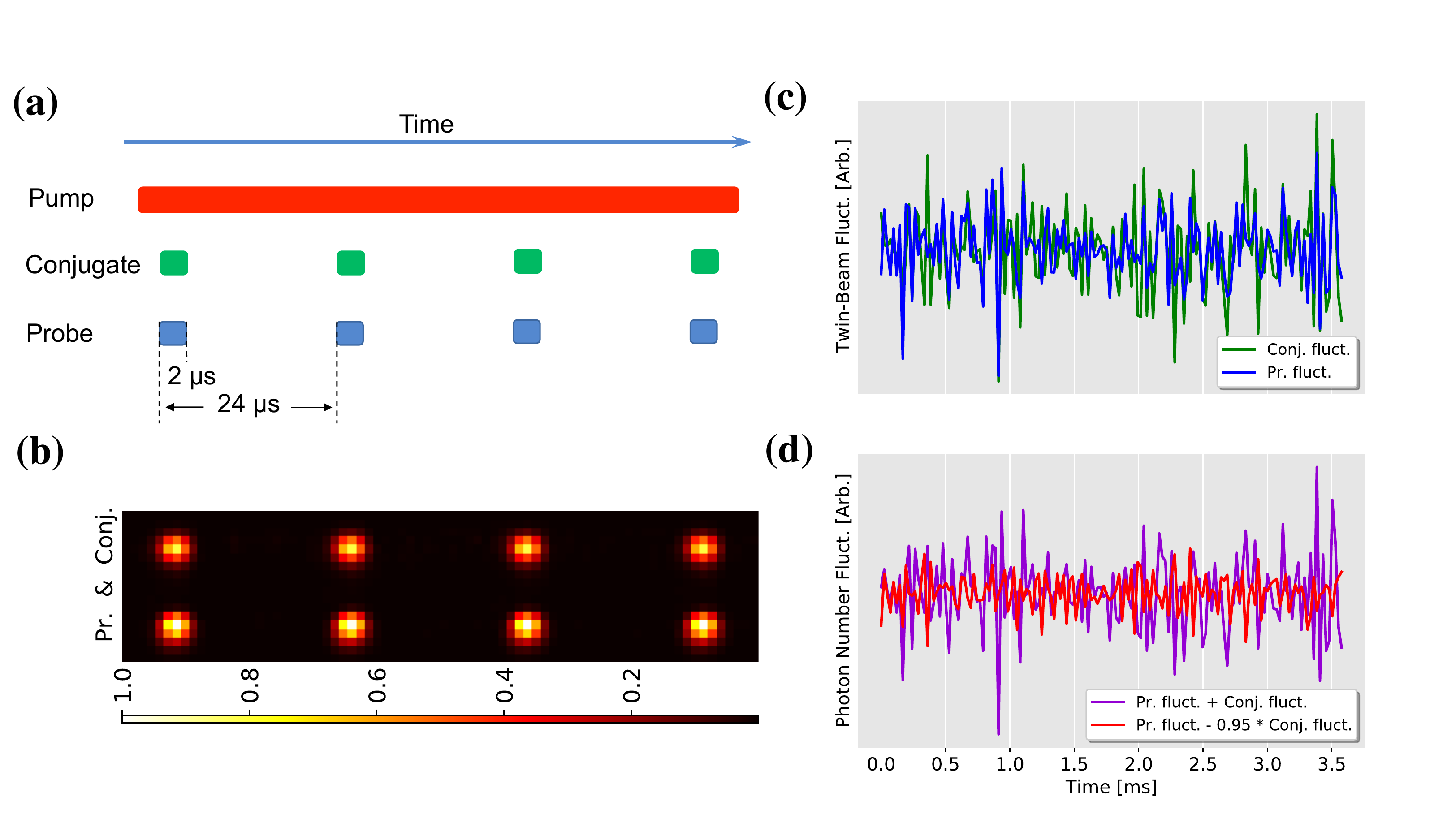}
    \caption{ \textcolor{black}{(a) Time sequencing of the pump and twin beams. The pulse duration of 2~$\mu$s and duty cycle of 1/12 is realized by pulsing the probe beam with an AOM. The CW pump beam is present all the time. (b) Typical images of the twin beams with absorption $\alpha=3~\%$ captured by the EMCCD camera. \textcolor{black}{This subfigure is the `real life' version of subfigure (a). It is an image of four consecutive pulses with the pulse width and duty cycle shown in subfigure (a).} (c) Temporal photon counts fluctuations of the probe $N_p(t)$ and conjugate $N_c(t)$ obtained by integrating the photon counts in the cropped regions in (b). Clear similarities can be observed between the twin beams. (d) The strong noise reduction in the subtraction as opposed to the summation of the $N_p(t)$ and $N_c(t)$ depicted in (c) showcases strong correlations between them.} 
    \label{Method}}
\end{figure*}

The experimental setup and the respective $^{85}$Rb atomic level structure are shown in Fig.~\ref{Setup}(a) and (b). The atomic medium is pumped by a strong ($\sim 500$~mW) narrow-band continuous-wave (CW) laser at frequency $\nu_1$ ($\lambda = 795$~nm) with a typical linewidth $\Delta \nu_1 \sim 100$~kHz. Applying an additional weak ($\sim$~10~nW) coherent seed beam 
at frequency $\nu_p = \nu_1 - (\nu_{HF}+\delta)$, where $\nu_{HF}$ and $\delta$ are the hyperfine splitting in the electronic ground state of $^{85}$Rb and the two-photon detuning respectively in Fig.~\ref{Setup}(b) (further experimental details can be found in Ref.~\cite{Li:20}), two pump photons are converted into a pair of twin photons, namely `probe $\nu_p$' and `conjugate $\nu_c$' photons, adhering to the energy conservation $2 \nu_1 = \nu_p + \nu_c$ (see the level structure in Fig.~\ref{Setup}(b)). The resulting twin beams are strongly quantum-correlated and are also referred to as bright two-mode squeezed light~\cite{PhysRevA.78.043816}. As can be seen from Fig.~\ref{Setup}(c), the twin beams exhibit a intensity-difference squeezing of 6.5~dB measured by balanced photodiodes (see Ref.~\cite{Li:20} for further details on the squeezing measurement), which is indicative of strong quantum correlations~\cite{PhysRevA.78.043816}.

After the $^{85}$Rb vapor cell,  the pump and the twin beams are separated by a second polarizer, with $\sim$~$2\times10^5$~$:1$ extinction ratio for the pump. The probe beam transverses through a combination of a $\lambda/2$ plate and a PBS, acting as an absorption sample, while the conjugate beam serves as a reference. The twin beams are then focused onto an EMCCD camera (Andor iXon Ultra 897). The EMCCD camera is enclosed in a light-proof box with filters installed at the entrance to block ambient light photons from entering the camera. The acousto-optic modulator (AOM) in the probe beam path is used to pulse the beam with 2~$\mu$s duration (FWHM) and duty cycle of $1/12$. Since the CW pump beam is present all the time, the conjugate beam is therefore also pulsed as a result of the FWM process.  The time sequencing of the pump and the twin beams are shown in Fig.~\ref{Method}(a) as the red strap, and the blue and green squares respectively.


We acquire the temporal behavior of the twin beams through the use of the \textit{kinetic} mode of the EMCCD camera. The EMCCD has $512\times512$ pixels with each pixel size of 16~$\mu$m$\times$16~$\mu$m. We focus the twin beams on the camera with an $1/e^2$ beam diameter of $\sim 50~\mu$m, occupying roughly 3 pixels as shown in Fig.~\ref{Method}(b). The temperature of the EMCCD is kept low ($<-65^\circ$C) to curb the thermal noise contributions. 
The rest of the EMCCD camera settings can be found in Ref.~\cite{Li:20}.




For each absorption $\alpha$ (acquired by changing the angle of the $\lambda/2$ plate), we capture 200 kinetic series, i.e., 200 frame sequences, with each frame having 35 pairs of probe and conjugate images containing the desired absorption information. We then crop a $10\times$10 pixel region around the maximum-intensity area in each probe and conjugate images, large enough to enclose their respective full beam profiles (see Fig.~\ref{Method}(b)), we thus can obtain the average total number of photons in the probe beam $\bar{N}_p$ and in the conjugate beam $\bar{N}_c$ by integrating photon counts in the cropped regions.

\textcolor{black}{The measurement signal $S_{\alpha}$ is defined as the photon number difference between the probe and conjugate beams:}

\begin{equation}
\begin{aligned} 
S_{\alpha} \equiv \bar{N}_{p} - \bar{N}_{c} = (1-\alpha)\bar{N}_{p0} - \bar{N}_{c},
\label{Sig}
\end{aligned} 
\end{equation} 
\textcolor{black}{where $\bar{N}_{p0}$ and $\bar{N}_p$ are the average numbers of photons in the probe beam before and after the faint absorber respectively, and $\bar{N}_{c}$ is the average number of photons in the conjugate beam. Factoring out $\alpha$, we have}

\begin{equation}
\begin{aligned} 
\alpha = -\frac{1}{\bar{N}_{p0}} S_{\alpha} + \frac{S_0}{\bar{N}_{p0}},
\label{Alpha}
\end{aligned} 
\end{equation} 
\textcolor{black}{where $S_0 \equiv \bar{N}_{p0} - \bar{N}_{c}$ is the photon number difference of the twin beams without the presence of the absorber, which can be treated as a characteristic of the quantum light source itself.}

\textcolor{black}{Also, the relation between the uncertainties of absorption $\alpha$ and the measurement signal $S_{\alpha}$ can be derived from the error propagation formula (see Eq.~(\ref{DEL})):}

\begin{equation}
\begin{aligned} 
\Delta \alpha= \frac{\Delta S_{\alpha} }{|\partial_\alpha {S_{\alpha}}|} = \frac{1}{\bar{N}_{p0}}\Delta S_{\alpha}, 
\label{Del}
\end{aligned} 
\end{equation}
\textcolor{black}{where $|\partial_\alpha {S_{\alpha}}| = \bar{N}_{p0}$ is obtained from Eq.~(\ref{Sig}).}
\textcolor{black}{Therefore following Eqs.~(\ref{Alpha}) and (\ref{Del}), the absorption $\alpha$ and its sensitivity $\Delta \alpha$ can be readily obtained from the measurements of $S_{\alpha}$ and $\Delta S_{\alpha}$.}

\begin{figure}[t]
    \begin{center}
    \includegraphics[width=\linewidth]{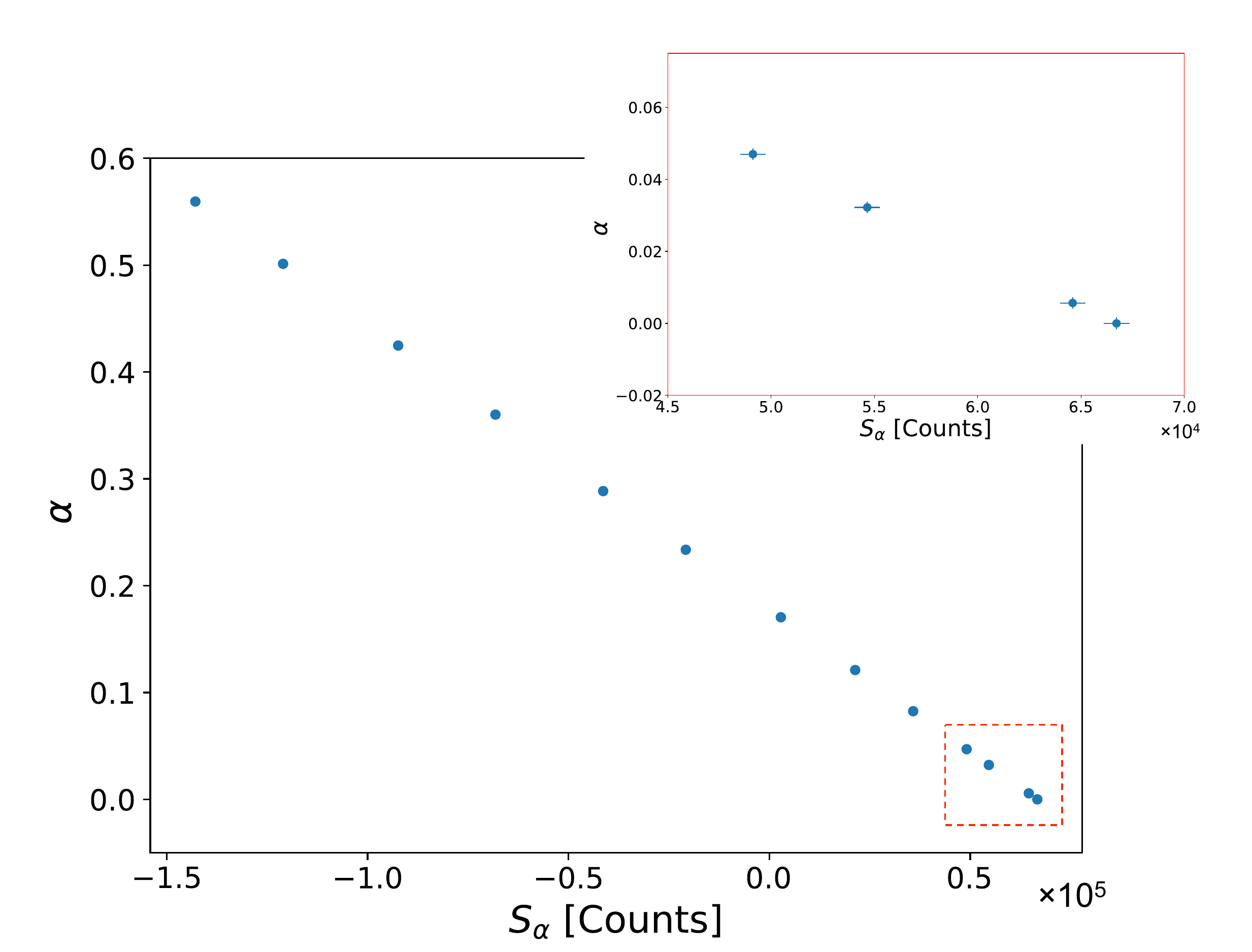}
    \caption{
    \textcolor{black}{Actual absorption $\alpha$  as a function of the measurement signal $S_{\alpha}$ defined in Eq.~(\ref{Sig}). The inset is a zoom-in view of \textcolor{black}{the data points with absorption $\alpha<10$~\%} to illustrate the sizes of \textcolor{black}{the uncertainties on the $x$-axis, $\Delta S_{\alpha}$, and the uncertainties on the $y$-axis, $\Delta {\alpha}$}.}
        \label{ActualMeasurement}}
    \end{center}
\end{figure}

\textcolor{black}{In Fig.~\ref{ActualMeasurement}, we plot the actual absorption $\alpha$ as a function of the measurement signal $S_{\alpha}$. The inset in Fig.~\ref{ActualMeasurement} is a zoom-in view of \textcolor{black}{the data points with absorption $\alpha<10$~\%} to illustrate the sizes of uncertainties of these two quantities, i.e., $\Delta S_{\alpha}$ on the $x$-axis and $\Delta \alpha$ on the $y$-axis.} \textcolor{black}{In the experiment,  we observe $1.3\pm 0.2$~dB quantum advantage in terms of $\Delta S_{\alpha}$ when comparing to shot-noise limited classical measurements for faint absorption levels (see Fig.~\ref{Del}). Due to the fact that $\Delta \alpha \propto \Delta S_{\alpha}$ with $1/\bar{N}_{p0}$ be the proportionality constant (see Eq.~(\ref{Del})), this greater than 1~dB quantum advantage should also translate to $\Delta \alpha$ when compared to its shot-noise limited classical counterparts.}

For measurements of  the quantum noise reduction between the twin beams, we adopt an algorithm originally developed in the spatial domain~\cite{PhysRevA.95.053849,PhysRevA.100.063828} but re-deriving it in the temporal domain. As shown in Fig.~\ref{Method}(c), the temporal photon counts fluctuations of the probe beam $N_p(t)$ and conjugate beam $N_c(t)$ are acquired by integrating photon counts in the cropped $10\times$10 pixel regions for 7000 pairs of twin-beam images during 170~ms.
As expected, strong correlations between the photon counts fluctuations of the twin beams can be observed in Fig.~\ref{Method}(c) and manifested in Fig.~\ref{Method}(d) through the subtraction and addition of these two modes. The quantum noise reduction characterization, $\sigma$, in the temporal domain reads

\begin{equation}
\begin{aligned}
\sigma\equiv\frac{\langle\Delta^2[(N_p(t+\delta t)-N_p(t))-\eta(N_c(t+\delta t)-N_c(t))]\rangle_t}{\langle N_p(t+\delta t)+ N_p(t) + \eta N_c(t+\delta t) + \eta N_c(t)\rangle_t},
\end{aligned}
\label{sigma}
\end{equation}


where $N_p(t+\delta t)-N_p(t)$ and $N_c(t+\delta t)-N_c(t)$ are the subtractions of photon counts in the cropped regions in two successive probe and conjugate images with time interval of $\delta t = 24$~$\mu$s. Since intensities of the twin beams are inherently imbalanced due to the seed power and different transmissions through the vapor cell~\cite{Li:17}, a scaling factor $\eta=0.95$, which is obtained by taking the ratio between the conjugate and probe photon counts in the analysis regions without the presence of the absorption sample, is applied to the conjugate mode to rescale its photon count before the two modes are subtracted. 
\textcolor{black}{Note that each image is involved in averaging over the spatial intensity profile of the beam, and the scaling factor effectively balances not only any differences in the averaging of the beam intensity profiles but also the intensity fluctuations. The subtraction of two successive images leads to the cancellation of the low-frequency portion of classical noise so that the rest of fluctuations are in the shot-noise-limited regime~\cite{PhysRevA.95.053849,PhysRevA.100.063828}.} The numerator of Eq.~(\ref{sigma}) represents the temporal variance of the intensity-difference noise between the probe and conjugate pulses. The denominator gives the mean photon counts for the probe and conjugate pulses used for the analysis and represents the shot noise. For coherent state pulses $\sigma =1$, which corresponds to the SNL, while for thermal light or other classical states $\sigma >1$.  Temporally quantum-correlated beams, like the twin beams generated in our experiment, will result in $\sigma <1$, with a smaller $\sigma$ corresponding to a larger degree of quantum correlations (i.e., two-mode squeezing).

\begin{figure}[t]
    \begin{center}
    \includegraphics[width=\linewidth]{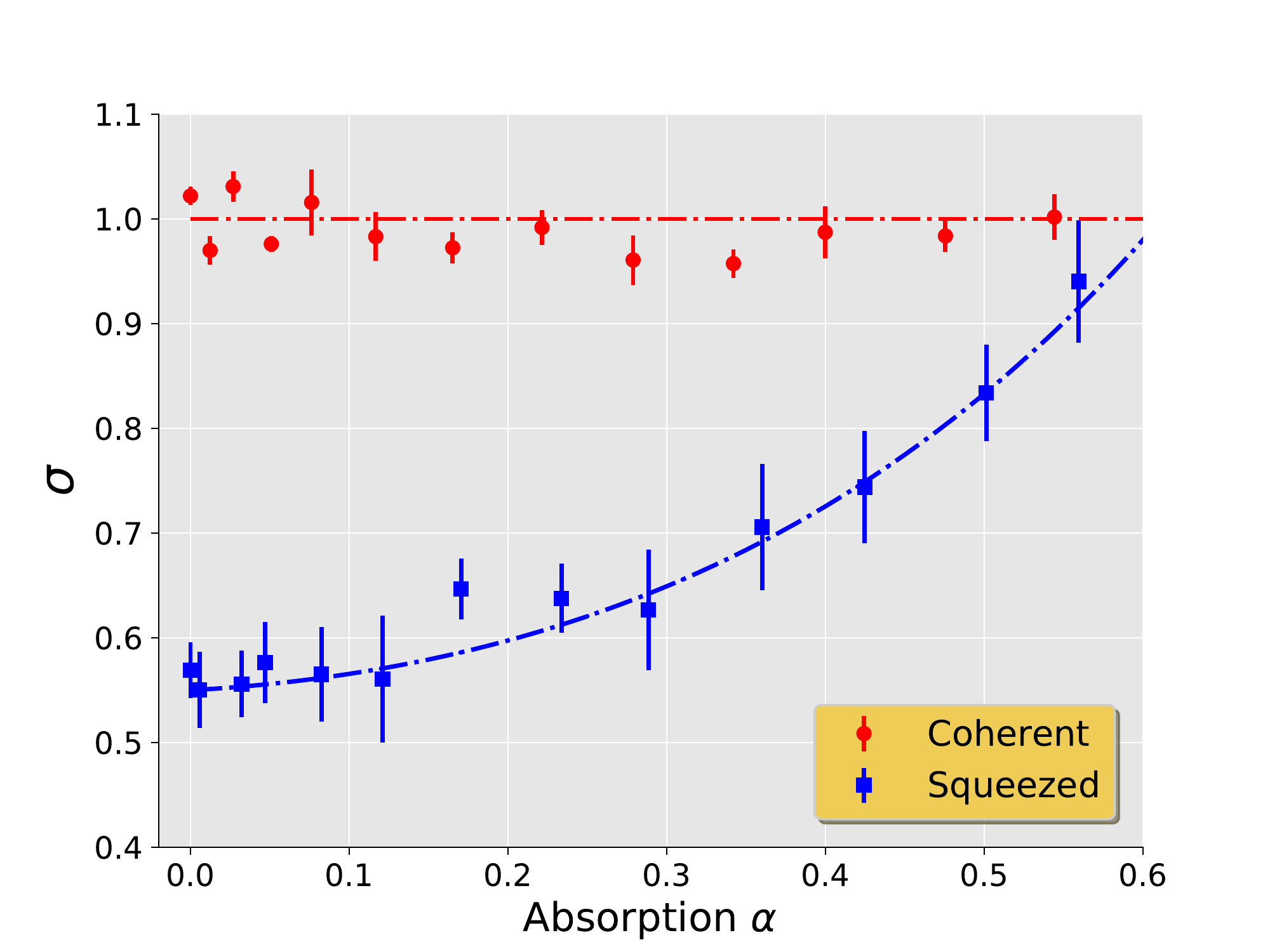}
    \caption{
    \textcolor{black}{Temporal quantum noise reduction $\sigma$ as a function of absorption $\alpha$ for the intensity squeezed light (blue squares) and coherent light (red dots). Dashed blue line is the theoretical prediction with $\eta_p=0.61$, $\eta_c=0.63$ and $r=1.1$.} 
        \label{Sigma}}
    \end{center}
\end{figure}

In Fig.~\ref{Sigma}, we plot $\sigma$ as a function of absorption $\alpha$ for the squeezed light together with coherent light. For each $\alpha$, we average 5 sets of 200 kinetic series and designate the error bar with one standard deviation. As expected, $\sigma < 1$ for the squeezed light (blue squares), while $\sigma \cong 1$ when the twin beams are replaced with two coherent beams (red dots). The notable degradation of the temporal quantum noise reduction measured by the EMCCD camera with respect to the one measured by balanced photodiodes in Fig.~\ref{Setup}(c) can be mainly attributed to 
a much worse quantum efficiency of the EMCCD camera at 795~nm (merely 70~\%  as opposed to at least 94~\% for photodiodes). We also repeated the experiment with different pulse duty cycles (i.e., $\delta t$ in Eq.~(\ref{sigma})), but they seemed to play an nonessential role on the quantum noise reduction as long as we were in the shot-noise-limited regime, i.e., $\sigma$ is still close to unity for coherent beams. 

\begin{figure}[t]
    \begin{center}
    \includegraphics[width=\linewidth]{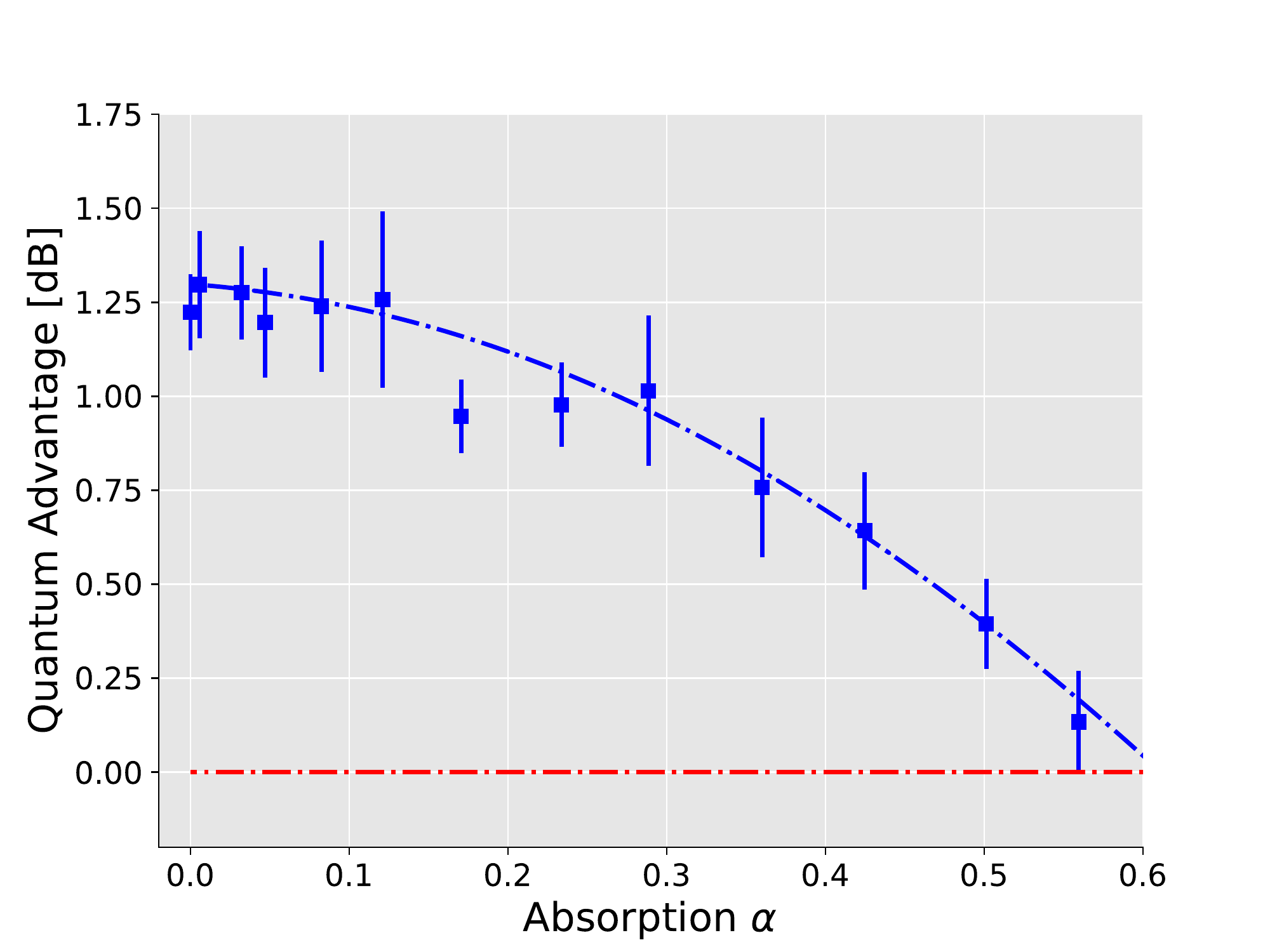}
    \caption{
    \textcolor{black}{Quantum advantage as a function of absorption $\alpha$. Dashed blue line is the theoretical prediction with $\eta_p=0.61$, $\eta_c=0.63$ and $r=1.1$. The quantum advantage is only significant ($> 1$~dB) for small values of $\alpha$ ($< 20~\%$), and for $\alpha > 60~\%$ there is no quantum advantage.}
        \label{QuantumAdvantage}}
    \end{center}
\end{figure}

From Eqs.~(\ref{DEL}) and~(\ref{QD}) we can easily arrive at

\begin{equation}
\begin{aligned} 
\begin{split}
\text{Quantum Advantage (dB)} = 10\times\text{log}_{10}\frac{\Delta \alpha_{\text{sqz}}}{\Delta \alpha_{\text{snl}}}\\
= 10\times\text{log}_{10}\sqrt{\frac{\langle\Delta^2 \hat{N_{\alpha}}\rangle_{\text{snl}}}{\langle\Delta^2 \hat{N_{\alpha}}\rangle_{\text{sqz}}}} = 10\times\text{log}_{10}\sqrt{\frac{1}{\sigma}}.
\end{split}
\label{QD2}
\end{aligned} 
\end{equation}
We thus can use the same data depicted in Fig.~\ref{Sigma} to plot the quantum advantage versus absorption $\alpha$. The results are shown in Fig.~\ref{QuantumAdvantage}. Theoretical predictions for the temporal quantum noise reduction characterization $\sigma$ and the quantum advantage as a function of absorption $\alpha$ are plotted as dashed blue lines in Figs.~\ref{Sigma} and~\ref{QuantumAdvantage}, where excellent agreements between experiment and theory can be seen. At those faint absorption levels ($\alpha \leq 10$~\%) in Fig.~\ref{QuantumAdvantage}, the observed quantum advantage can be more than 1.2 dB, although \textcolor{black}{\textit{total}} optical losses \textcolor{black}{(including the transmission loss imposed by optics and imperfect detection quantum efficiency imposed by the EMCCD camera)} in the paths of the twin beams are significant - nearly $39$~\% in the probe path and nearly $37$~\% in the conjugate path. This is mainly due to a relatively low quantum efficiency of the EMCCD camera at 795~nm ($\sim 70~\%$) and imperfect transmission of the band pass filters ($\sim94\%$) mounted in front of the light-proof box. \textcolor{black}{If we were able to overcome this main obstacle of the experiment by employing a much more efficient camera, we would have a much higher quantum advantage approaching 3~dB as implied by the theoretical curves in Fig.~\ref{Theory}. We notice that a most recent work~\cite{doi:10.1063/5.0009681} has also demonstrated quantum advantage in absorption measurement using a single-mode amplitude squeezed light generated with an optical phase-sensitive amplifier. The reported advantage (according to Fig.~5 in Ref.~\cite{doi:10.1063/5.0009681}) is less than 0.5~dB.}

\textcolor{black}{It is worth mentioning that taking measurements using photodetectors would yield better results due to photodiodes' much higher quantum efficiency. However, the main drawback of using photodetectors is their much higher power requirement. For an EMCCD camera,  a few nW input power is more than enough to yield a clear signal-to-noise ratio, however, for a photodetector to provide sufficient signal clearance from its electronic noise floor, the input power has to be in the range of tens of $\mu$W. For example in our experiment, in order to have a signal noise power that is 10 dB above the electronic noise floor, we have to shine a coherent beam of light of at least 50~$\mu$W to the photodetector (given our squeezing level of 6.5~dB, that implies a merely 3.5 dB clearance from the electronic noise floor for 50~$\mu$W squeezed light). One of the most important implementations of our experimental scheme is to characterize biological samples without imposing light-induced damages, a much higher input light power would hence defeat this purpose.}\\


\section{Discussion}

Overall, our experiment realizes a practical scheme that allows the SNL in the direct absorption measurement to be overcome. We demonstrate that by using the intensity squeezed light more than 1.2~dB quantum advantage is achieved for the measurement sensitivity at faint absorption levels ($\leq10\%$). We thus experimentally demonstrate the advantage of squeezed light for measurements on open systems. \textcolor{black}{We also theoretically demonstrate that more quantum advantage ($>$~3~dB) is very likely attainable by means of a proper optical loss management.} We use seeded FWM process in an atomic $^{85}$Rb vapor cell to generate the quantum-correlated twin beams of light. It is also the first experiment that uses quantum light generated with FWM instead of SPDC to demonstrate a sub-shot-noise absorption measurement. Major advantages of this FWM-based quantum light generation scheme include an ultra-high photon-pair flux up to~$10^{16}$ photons/s, which is a few orders of magnitude higher than the fluxes produced by SPDCs~\cite{Jechow:08,Villabona-Monsalve:2018fy,Varnavski:2017rp}, and narrow-band probe and conjugate beams ($\sim 20$~MHz)~\cite{Clark:2014vf,Glasser2012a}, which can be readily integrated into quantum networks through coupling with micro-resonators/cavities. \textcolor{black}{Also, although the small bandwidth feature of the twin beams is not used in the experiment, we do take advantage of it by making a `single-mode' approximation for the twin beams in the theoretical analysis. The fact that our experimental results agree very well with the theory based on the `single-mode' approximation confirms the importance of the narrow band feature of the twin beams.} Moreover, the seeded FWM process offers sufficient gains in a single-pass configuration producing bright quantum-correlated beams of light without a cavity, making it possible to preserve the multi-spatial-mode nature of the bright twin beams~\cite{PhysRevLett.109.043602,Corzo:11}. Our quantum light generation together with the direct absorption measurement scheme reported here can be therefore greatly beneficial to many applications involving characterizing chemical and biological samples, where sub-SNL absorption measurements are highly desirable~\cite{Genovese_2016,QuantumImaging}. 


We gratefully acknowledge the support of Air Force Office of Scientific Research (Award No. FA-9550-18-1-0141), Office of Naval Research (Award No. N00014-20-1-2184), and the Robert A. Welch Foundation (Grants No. A-1261 \& A-1943). F.L. acknowledges support from the Herman F. Heep and Minnie Belle Heep Texas A\&M University Endowed Fund held/administered by the Texas A\&M Foundation.


\bibliography{MyLibrary.bib}

\end{document}